\documentclass{appolb}
\usepackage{calc}

\usepackage{graphicx}
\usepackage{dcolumn}
\usepackage{bm}
\usepackage{slashed}
\usepackage{amsmath,graphicx}
\usepackage[colorlinks=true,linktocpage=true,linkcolor=blue,citecolor=blue]{hyperref}
\usepackage{float}
\usepackage{nicefrac}
\usepackage[normalem]{ulem}
\usepackage{amsmath}
\usepackage{subfigure}

\usepackage{float} 
\usepackage{bbold}

\newcommand{\beq}{\begin{eqnarray}}
\newcommand{\eeq}{\end{eqnarray}}

\newcommand{\bel}[1]{\begin{eqnarray}\label{#1}}
\newcommand{\eel}{\end{eqnarray}}

\newcommand{\rf}[1]{Eq.~(\ref{#1})}

\newcommand{\rfn}[1]{(\ref{#1})}

\newcommand{\rff}[1]{Fig.~\ref{#1}}

\newcommand{\nn}{\nonumber}

\newcommand{\p}{\partial}

\newcommand{\trf}{{\rm tr_4}}

\newcommand{\f}[2]{\frac{#1}{#2}}
\newcommand{\onehalf}{{\nicefrac{1}{2}}}

\renewcommand\sout{\bgroup \color{blue} \ULdepth=-.5ex \ULset}

\newcommand{\ed}{{\varepsilon}}       

\def\gmnU{g^{\mu\nu}}

\def\TmnU{T^{\mu \nu}}

\def\n0{n_{(0)}}
\def\e0{\varepsilon_{(0)}}
\def\P0{P_{(0)}}
\def\s0{s_{(0)}}

\def\fplusrsxp{f^+_{rs}(x,p)}

\def\fminusrsxp{f^-_{rs}(x,p)}

\def\bmu{\beta_\mu}

\def\umU{u^\mu}  
\def\unU{u^\nu}

\def\omnL{\omega_{\mu\nu}}
\def\omnU{\omega^{\mu\nu}}

\def\pmu{p^\mu}
\def\pnu{p^\nu}

\def\omU{\omega^\mu}

\def\SmunuU{{\Sigma}^{\mu\nu}}

\def\S0iU{{\Sigma}^{0i}}

\def\ubarrp{{\bar u}_r(p)}

\def\usp{u_s(p)}
\def\urp{u_r(p)}

\def\vbarsp{{\bar v}_s(p)}

\def\vrp{v_r(p)}

\def\g5{\gamma_5}

\def\slmnU{S^{\lambda, \mu \nu}}

\def\Ot{\tilde \Omega}

\newcommand{\lsb}{\left[}
\newcommand{\rsb}{\right]}

\begin{document}
\title{
Fluid dynamics for relativistic spin-polarized media
\thanks{Presented by Radoslaw Ryblewski at
Excited QCD 2018, March 11-15, 2018, Kopaonik, Serbia.}%
}
\author{Wojciech Florkowski
\address{Institute of Nuclear Physics, PL-31342 Krak\'ow, Poland \\
	Jan Kochanowski University, PL-25406 Kielce, Poland }\\
	 \vspace{0.3cm}
	Bengt Friman
		\address{GSI Helmholtzzentrum f\"{u}r Schwerionenforschung, D-64291 Darmstadt, Germany}	\\
	 \vspace{0.3cm}
		Amaresh Jaiswal
		\address{ 
			School of Physical Sciences, National Institute of Science Education and
			Research, HBNI, Jatni-752050, India}\\
	 \vspace{0.3cm}
		 Radoslaw Ryblewski
		 \address{ 
		 	Institute of Nuclear Physics, PL-31342 Krak\'ow, Poland}
		 \and
			Enrico Speranza
			\address{
				Institute for Theoretical Physics, Goethe University,\\
				D-60-438 Frankfurt am Main, Germany}
}
\maketitle
\begin{abstract}
We briefly review the basic features of a new framework for relativistic perfect fluid hydrodynamics of polarized systems consisting of particles with spin one half. Using this approach we numerically study the stability of a stationary vortex-like solution, representing global equilibrium of a rotating medium.
\end{abstract}
%
\section{Introduction}
%
The recent observation of global spin polarization of $\Lambda$ hyperons by the STAR collaboration \cite{STAR:2017ckg} has rekindled the interest in polarization and vorticity in ultrarelativistic heavy-ion collisions. In contrast to a multitude of classical effects \cite{Florkowski:2010zz,Florkowski:2017olj}, the polarization of spin represents a first, rather clear, experimental manifestation of a pure quantum phenomenon in nucleus-nucleus collisions. A particularly appealing theoretical explanation of this effect invokes a direct coupling between the thermal vorticity and polarization, which is realized in the global equilibrium state of a rotating medium~\cite{Becattini:2009wh,Becattini:2013fla}. Recently, a new framework for relativistic perfect fluid hydrodynamics of spin-polarized media was presented~\cite{Florkowski:2017ruc,Florkowski:2017njj,Florkowski:2017dyn,Florkowski:2018myy}, which extends the work of Refs.~\cite{Becattini:2009wh,Becattini:2013fla} to systems in local equilibrium. In this contribution we discuss the approach presented in Refs.~\cite{Florkowski:2017ruc,Florkowski:2017njj,Florkowski:2017dyn,Florkowski:2018myy} and study its physical consequences.
%
\section{Local equilibrium distribution functions}
%
Following Refs.~\cite{Florkowski:2017ruc,Florkowski:2017njj,Florkowski:2017dyn,Florkowski:2018myy} we consider a  local equilibrium state of a relativistic system of particles ($+$) and antiparticles ($-$) with spin $\onehalf$ and mass $m$,  whose phase-space distribution functions are given by the spin density matrices ($r,s = 1,2$)   \cite{Becattini:2013fla}
\bel{fplusrsxp}
\fplusrsxp =  \ubarrp X^+ \usp, \qquad
\fminusrsxp = - \vbarsp X^- \vrp.
\eel
Here $\urp$ and $\vrp$ are Dirac bispinors and $X^{\pm}$ are  four-by-four matrices 
\bel{XpmM}
X^{\pm} =  \exp\left[\pm \xi(x) - \bmu(x) \pmu  \pm \f{1}{2} \omnL(x)  \SmunuU \right], 
\eel
where  $\xi \equiv \mu/T$, $\beta^\mu \equiv \umU/T$, and $T$, $\mu$ and $\umU$ denote the temperature, baryon chemical potential and four-velocity of the fluid, respectively. The quantity $\omnL$ is the spin polarization tensor and $\SmunuU  \equiv \f{1}{2} \sigma^{\mu\nu} = \f{i}{4} [\gamma^\mu,\gamma^\nu]$ is the spin operator. For later convenience it is useful to define the quantity $\zeta  \equiv \f{1}{2 \sqrt{2}} \sqrt{ \omnL \omnU }$, assuming that $\epsilon_{\alpha\beta\gamma\delta} \omega^{\alpha\beta}\omega^{\gamma\delta}=0$.
%
\section{Thermodynamics of the spin-polarized medium}
%
With the distribution functions \rfn{fplusrsxp} being defined it is straightforward to obtain the basic thermodynamic variables describing the system in question. In particular, using definitions from  Refs.~\cite{Becattini:2013fla,DeGroot:1980dk}, one finds the following familiar expressions for the baryon current and the energy-momentum tensor, 
\beq 
N^\mu &=&  \kappa \int \f{d^3p}{2 E_p}  \pmu \left[ \trf( X^+ ) - \trf ( X^- )  \right] = n \umU, \label{Nmu}\\
\TmnU &=&  \kappa \int \f{d^3p}{2 E_p}  \pmu \pnu  \left[ \trf( X^+ ) + \trf ( X^- )  \right]   \label{Tmunu} = (\ed + P) \umU \unU - P \gmnU.
\eeq
In Eqs.~(\ref{Nmu})--(\ref{Tmunu}) the factor $\kappa \equiv g/(2\pi)^3$ accounts for internal degrees of freedom excluding spin, $\trf$ denotes the trace in the spinor space, and $\gmnU={\rm diag}(+1,-1,-1,-1)$ is the metric tensor. Generalizing the Boltzmann expression for the entropy current one also finds 
\beq 
S^\mu =   -\kappa \int \f{d^3p}{2 E_p}  \pmu \left\{ \trf[   X^+ (\ln X^+ -1) ] + \trf [  X^- (\ln X^- -1)]  \right\} = s \umU \label{Smu}. 
\eeq
In Eqs.~(\ref{Nmu})--(\ref{Smu}) the energy density, $\ed = 4  \cosh(\zeta)  \cosh(\xi)  \e0(T)$, the pressure, $P = 4  \cosh(\zeta)  \cosh(\xi) \P0(T)$,  the baryon density, $n = 4 \cosh(\zeta)   \sinh(\xi)  \n0(T)$, and the entropy density, $s = 4  \cosh(\zeta)  \cosh(\xi) \s0(T)$, are all related by the thermodynamic relation $\ed + P= s T +\mu n + \Omega w$. Here we introduced a new variable, playing the role of a spin chemical potential $\Omega \equiv \zeta T$, and, related to it,  a new charge, $w = 4  \sinh(\zeta) \cosh(\xi) \n0(T)$. The thermodynamic potentials are expressed in terms of those corresponding to an auxiliary system of spin-$0$ particles
\beq
\n0(T) &=& \langle(u\cdot p)\rangle_0 = \f{\kappa}{2\pi^2}\, T^3 \, \hat{m}^2 K_2\left( \hat{m}\right) \,, \label{polden}\nn\\
\e0(T) &=& \langle(u\cdot p)^2\rangle_0= \f{\kappa}{2\pi^2} \, T^4 \, \hat{m} ^2
\Big[ 3 K_{2}\left( \hat{m} \right) + \hat{m}  K_{1} \left( \hat{m}  \right) \Big]\,,  \label{eneden}\nn\\
\P0(T) &=& -\f{1}{3} \langle \left[ p\cdot p - 
(u\cdot p)^2 \right] \rangle_0 =T \, \n0(T)  \,, \label{P0}\nn
\eeq
where $\s0(T) = \f{1}{T} \lsb\e0(T)+\P0(T)\rsb $ and $\hat{m}\equiv m/T$.
%
\section{Fluid dynamics equations}
%
In fluid dynamics the space-time dependent quantities $\mu(x)$, $T(x)$, $u^\mu(x)$ and $\omU$(x) in  \rf{XpmM} play a role of Lagrange multipliers, whose form should follow from the evolution equations. In particular, the requirement of energy and momentum conservation, $\p_\mu \TmnU = 0$, when projected onto directions orthogonal to the fluid flow, yields the three  Euler equations 
\beq
(\ed + P)  \dot{u}^\mu &=& \p^\mu P - \umU \dot{P}\label{emcprojt}\,,
\eeq
where  $\theta\equiv \p\cdot u$ is the expansion scalar and $\dot{(\hphantom{A})}\equiv u \cdot\p$ denotes the comoving derivative. On the other hand, by projecting the energy and momentum conservation equation onto the fluid four velocity $u^\mu(x)$, and using the differentials of the pressure $P=P(T,\mu,\Omega)$, one obtains
\bel{emcuproj}
T \p_\mu (s \umU)+\mu\, \p_\mu (n \umU)+\Omega\, \p_\mu (w \umU)=0. 
\eel
By requiring that the first two terms vanish due to the conservation of entropy and baryon number, respectively, \rf{emcuproj} yields three separate conditions
\beq
\p_\mu S^\mu=\dot{s} + s\, \theta &=&0 \label{entcons}\,, \\
\p_\mu N^\mu=\dot{n} + n\, \theta&=&0\label{chargcons}\,, \\
\p_\mu W^\mu =\p_\mu (w \umU)=\dot{w} + w\, \theta&=&0\label{polcons}\,.
\eeq
%
%
\section{Polarization dynamics}
%
Equations \rfn{emcprojt}, \rfn{entcons}, \rfn{chargcons} and \rfn{polcons} form a closed set of six differential equations, which allow one to determine time evolution of $\mu(x)$, $T(x)$, the three independent components of $u^\mu(x)$ and $\Omega(x)  \equiv \f{ T}{2 \sqrt{2}} \sqrt{ \omnL \omnU }$. We are thus left with four independent components of the polarization tensor, whose evolution does not influence the hydrodynamic background. The equations of motion for the polarization tensor follow from the angular momentum conservation law $\p_\alpha J^{\alpha,\beta\gamma} = 0$, with $J^{\alpha,\beta\gamma} = L^{\alpha,\beta\gamma} + S^{\alpha,\beta\gamma}$   where $L^{\alpha,\beta\gamma} = x^{\beta} T^{\gamma\alpha} - x^{\gamma} T^{\beta\alpha}$ is the orbital angular momentum tensor and $S^{\alpha,\beta\gamma}$ is the spin tensor. Since $T^{\mu\nu}$ given in \rf{Tmunu} is symmetric one has~\cite{Hehl:1976vr}
\beq
\p_\alpha S^{\alpha,\beta\gamma} = 0.
\eeq
For the internal consistency of the approach we assume that the spin tensor has the following form \cite{Becattini:2009wh}
\bel{st110}
\slmnU = \kappa \int \f{d^3p}{2 E_p} \, p^\lambda \, {\trf} \left[(X^+\!-\!X^-) \SmunuU \right] = \frac{w u^\lambda}{4 \zeta} \omega^{\mu\nu}.
\eel
By introducing the rescaled spin polarization tensor, $\bar{\omega}^{\mu\nu}=\omega^{\mu\nu}/(2\zeta)$ and using \rf{polcons}, one arrives at the formula
\bel{st}
\dot{\bar{\omega}}^{\mu\nu}   &=&0 \label{ptc} .  
\eel
Equation  \rfn{ptc} states that the scaled components of the polarization tensor are conserved in the comoving frame. 
%
\section{Stability of the stationary vortex solution}
%
In Refs.~\cite{Florkowski:2017ruc,Florkowski:2017dyn} it was shown that Eqs.~\rfn{emcprojt}, \rfn{entcons}, \rfn{chargcons} and \rfn{polcons} have the following stationary vortex-like solution,  representing a global equilibrium state with rotation \cite{Becattini:2009wh,Becattini:2013fla}
\bel{uvortex}
u^\mu  
&=& \gamma \,(1,  -  \Ot \, y,    \, \Ot \, x,  0), 
\eel
\vspace{-0.75cm}
\bel{Tvortex}
T &=& T_0 \gamma,  \quad \mu = \mu_0 \gamma,  \quad  \Omega = \Omega_0 \gamma,
\eel
where  $\gamma = 1/\sqrt{1 - \Ot^2 r^2}$ is the Lorentz factor,   $r = \sqrt{x^2 + y^2}$  and $T_0$, $\mu_0$, and $\Omega_0$ are arbitrary constants. The corresponding polarization tensor is in this case either zero or has a form where the only non-vanishing component is  ${  \omega}_{xy}=-{  \omega}_{yx} = \Ot/T_0  = 2 \, \Omega_0 /T_0$.
 %
 \begin{figure}[t]
 	\includegraphics[angle=0,width=0.45 \textwidth]{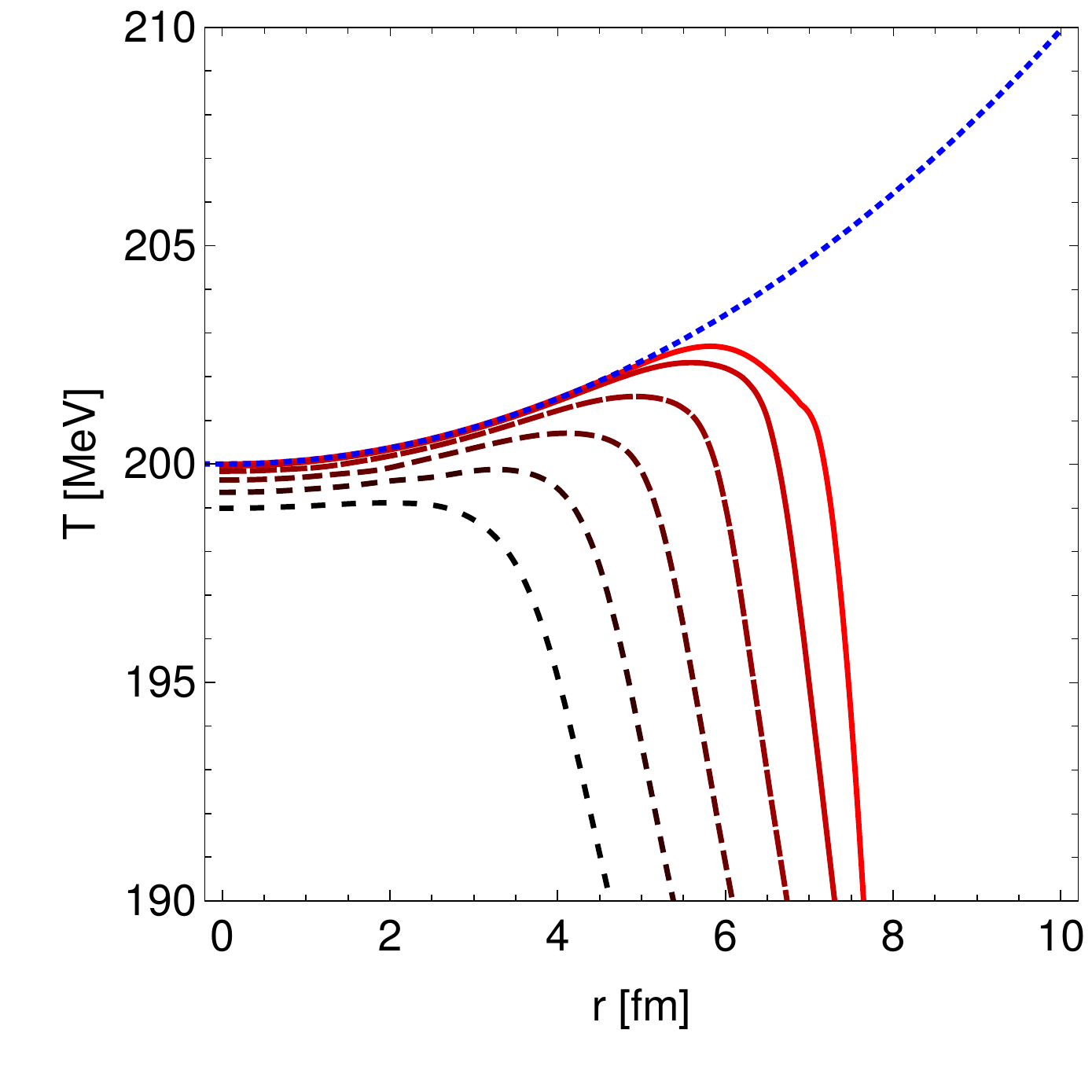}   
 	\includegraphics[angle=0,width=0.45 \textwidth]{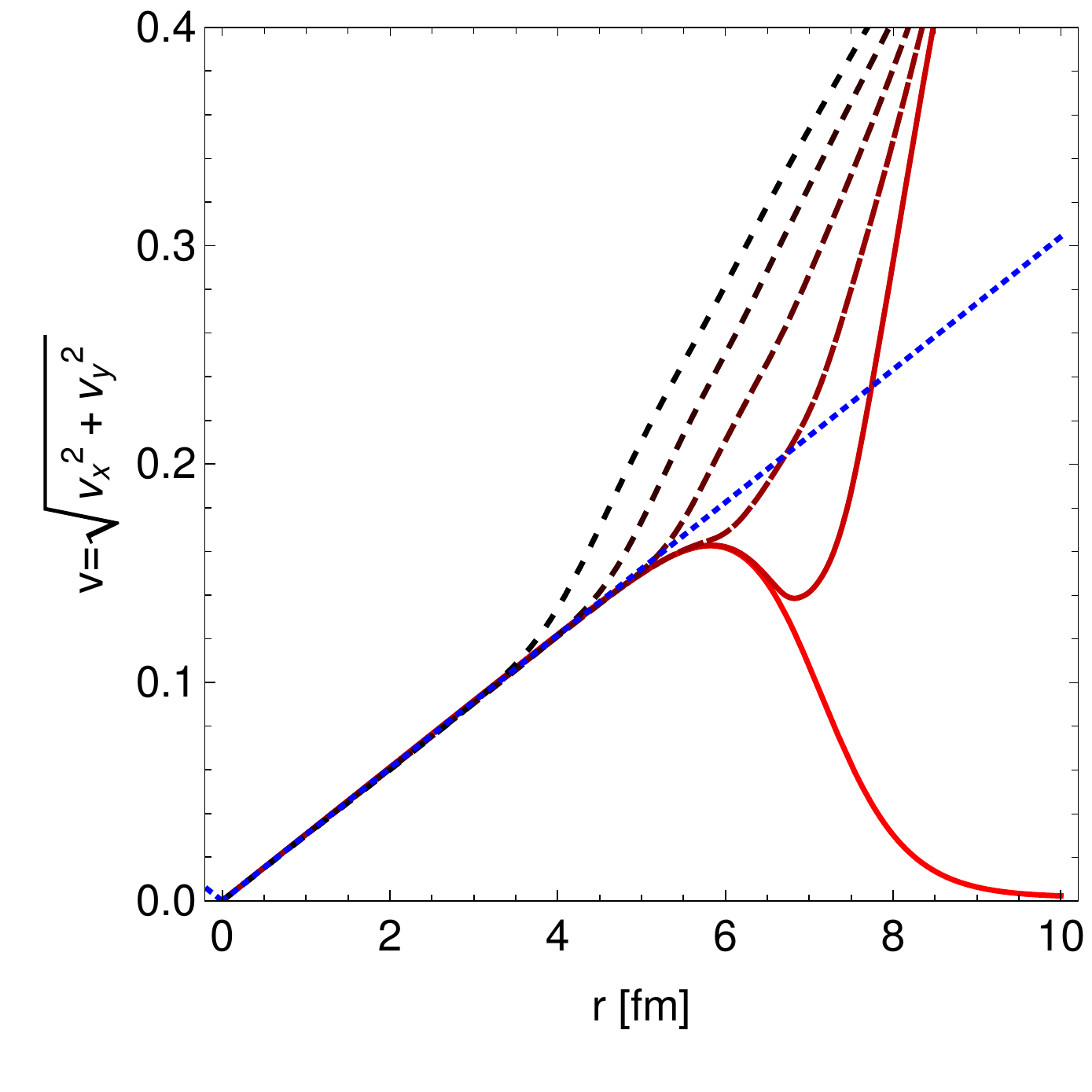} 
 	\caption{(Color online) Space ($r$) dependence of the temperature,  and velocity magnitude $v=\sqrt{v_x^2+v_y^2}$ at times: $t=0.1, 2, 4, 6, 8, 10$ fm (color of the lines changing from red to black as the time increases).
 	}
 	\label{fig:sol}
 \end{figure}
 %
 
One may notice that, due to the limiting speed of light, the stationary solution may be realized only within a cylinder of a finite  radius $R < 1/\Ot$. It is difficult to imagine how a corresponding boundary condition could be implemented in Nature.  Using the approach presented above, we explore the evolution of such a vortex, with more realistic boundary conditions numerically. To that end, we set up the initial conditions for the system in such a way that it reproduces the stationary vortex solution (dotted blue lines) within a central region (small $r$), and departs from it at the edges, as illustrated in \rff{fig:sol} (red lines). By letting the system evolve in time, we find that relaxing the boundary conditions causes the system to depart from the global equilibrium solution. Thus, our results indicate that, with realistic boundary conditions, the vortex solution is unstable.  

%
\section{Summary}
%
Using the framework of perfect fluid hydrodynamics of particles with spin $\onehalf$ we studied the stability of the stationary vortex-like solution representing the global equilibrium state of such a system. We find that with more realistic boundary conditions the stationary solution is unstable. Consequently, it is rather unlikely that such a stationary state is realized in Nature. 
%
\section*{Acknowledgments}
%
W.F. and R.R were supported in part by the Polish National Science Center Grant No.   2016/23/B/ST2/00717.  E.S.  was  supported  by  BMBF  Verbundprojekt 05P2015 - Alice at High Rate.  E.S. acknowledges partial support by the  Deutsche  Forschungsgemeinschaft  (DFG)  through  the  grant  CRC-TR 211 “Strong-interaction matter under extreme conditions”. A.J. is supported in part by the DST-INSPIRE faculty award under Grant No. DST/INSPIRE/04/2017/000038. This research was supported in part by the ExtreMe Matter Institute EMMI at GSI and was performed in the framework of COST Action CA15213 “Theory of hot matter and relativistic heavy-ion collisions” (THOR).
%

\end{document}